\documentclass[10pt, conference]{IEEEtran}
\IEEEoverridecommandlockouts

\usepackage{cite}
\usepackage{amsmath,amssymb,amsfonts}
\usepackage{algorithmic}
\usepackage{algorithm}
\usepackage{graphicx}
\usepackage{textcomp}
\usepackage{xcolor}
\usepackage{color}
\usepackage{booktabs}
\usepackage{tabularx}
\usepackage{multirow}
\usepackage{array}
\usepackage{adjustbox}
\usepackage{makecell}
\usepackage{float} 
\usepackage{balance}
\usepackage[most]{tcolorbox}
\def\BibTeX{{\rm B\kern-.05em{\sc i\kern-.025em b}\kern-.08em
    T\kern-.1667em\lower.7ex\hbox{E}\kern-.125emX}}

\newcommand{\boxmargin}{1mm}
\newtcolorbox{myboxc}{
    colback=gray!15!white,
    arc = 0pt, outer arc = 0pt,
    boxsep=0pt, left = 3pt, right = 0pt, top = 0pt, bottom = 0pt, 
    leftrule=3pt, bottomrule=0pt,toprule=0pt, rightrule=0pt,
    left = \boxmargin, right = \boxmargin, top = \boxmargin, bottom = \boxmargin
}

\begin{document}

\title{
SolContractEval: A Benchmark for Evaluating Contract-Level Solidity Code Generation

}

\author{
\IEEEauthorblockN{
Zhifan Ye\IEEEauthorrefmark{2}\IEEEauthorrefmark{4}, 
Jiachi Chen\IEEEauthorrefmark{2},
Zhenzhe Shao\IEEEauthorrefmark{3}, 
Lingfeng Bao\IEEEauthorrefmark{2}\IEEEauthorrefmark{4}, 
Xiaohu Yang\IEEEauthorrefmark{2}\IEEEauthorrefmark{4}, 
Zhongxin Liu\IEEEauthorrefmark{2}\IEEEauthorrefmark{4}\IEEEauthorrefmark{1}\thanks{\IEEEauthorrefmark{1}Corresponding author.}}

\IEEEauthorblockA{\IEEEauthorrefmark{2}The State Key Laboratory of Blockchain and Data Security, Zhejiang University, China\\
\IEEEauthorrefmark{4}Hangzhou High-Tech Zone (Binjiang) Institute of Blockchain and Data Security, China\\
\IEEEauthorrefmark{3}Sun Yat-sen University, China\\
\{yezhifan,lingfengbao,yangxh,liu\_zx\}@zju.edu.cn, chenjch86@mail.sysu.edu.cn, shaozhzh3@mail2.sysu.edu.cn}
}

\maketitle

\begin{abstract}

The rise of blockchain has brought smart contracts into mainstream use, creating a demand for smart contract generation tools. While large language models (LLMs) excel at generating code in general-purpose languages, their effectiveness on Solidity, the primary language for smart contracts, remains underexplored. Solidity constitutes only a small portion of typical LLM training data and differs from general-purpose languages in its version-sensitive syntax and limited flexibility. These factors raise concerns about the reliability of existing LLMs for Solidity code generation. Critically, existing evaluations, focused on isolated functions and synthetic inputs, fall short of assessing models' capabilities in real-world contract development.

To bridge this gap, we introduce \textbf{\textit{SolContractEval}}, the first contract-level benchmark for Solidity code generation. It comprises 124 tasks drawn from real on-chain contracts across nine major domains. Each task input, consisting of complete context dependencies, a structured contract framework, and a concise task prompt, is independently annotated and cross-validated by experienced developers. To enable precise and automated evaluation of functional correctness, we also develop a dynamic evaluation framework based on historical transaction replay.
Building on \textbf{\textit{SolContractEval}}, we perform a systematic evaluation of six mainstream LLMs. We find that Claude-3.7-Sonnet achieves the highest overall performance, though evaluated models underperform relative to their capabilities on class-level generation tasks in general-purpose programming languages.
Second, current models perform better on tasks that follow standard patterns but struggle with complex logic and inter-contract dependencies. Finally, they exhibit limited understanding of Solidity-specific features and contextual dependencies. 

\end{abstract}


\section{Introduction}

The ongoing advancement of blockchain technology and decentralized finance (DeFi) has led to the widespread adoption of smart contracts across various sectors, including finance, gaming, and supply chain~\cite{zheng2020overview}. According to DeFiLlama~\cite{defillama2025}, the total value locked (TVL) in Ethereum DeFi protocols exceeded \$80 billion by December 2024, marking a two-year high. Meanwhile, Etherscan~\cite{etherscan2025} data show a steady rise in the daily deployment of new smart contracts, reflecting growing demand for smart contract development. This surge underscores the need for smart contract development tools, including automated code generation solutions.

Recent advances in large language models (LLMs) have shown state-of-the-art performance in automated code generation, particularly in general-purpose languages such as Python and Java~\cite{wang2023review, jiang2024survey, deng2025enhancing}. OpenAI o1~\cite{OpenAIO1Hub}, for instance, achieves an impressive 89\% pass rate on the HumanEval benchmark~\cite{chen2021evaluating}. However, their effectiveness remains largely unverified for Solidity~\cite{solidity2025}, the primary language of blockchain development.
Moreover, Solidity accounts for only 0.01\% of The Stack v2, the pretraining dataset used by models like StarCoder 2~\cite{lozhkov2024starcoder}, indicating limited training exposure. Solidity differs fundamentally from general-purpose languages in both design and constraints, posing unique challenges for code generation. On one hand, the rapid evolution of the Solidity compiler introduces challenges such as frequent version updates, syntax changes, and inconsistent feature support. On the other hand, its limited language flexibility increases the difficulty of implementing complex functionalities, such as managing dynamic data structures or expressing advanced control logic. This naturally raises a critical question: given the scarcity of training data, can existing LLMs meet Solidity smart contract development requirements?

\begin{figure}[ht]
    \centering
    \includegraphics[width=\columnwidth, clip]{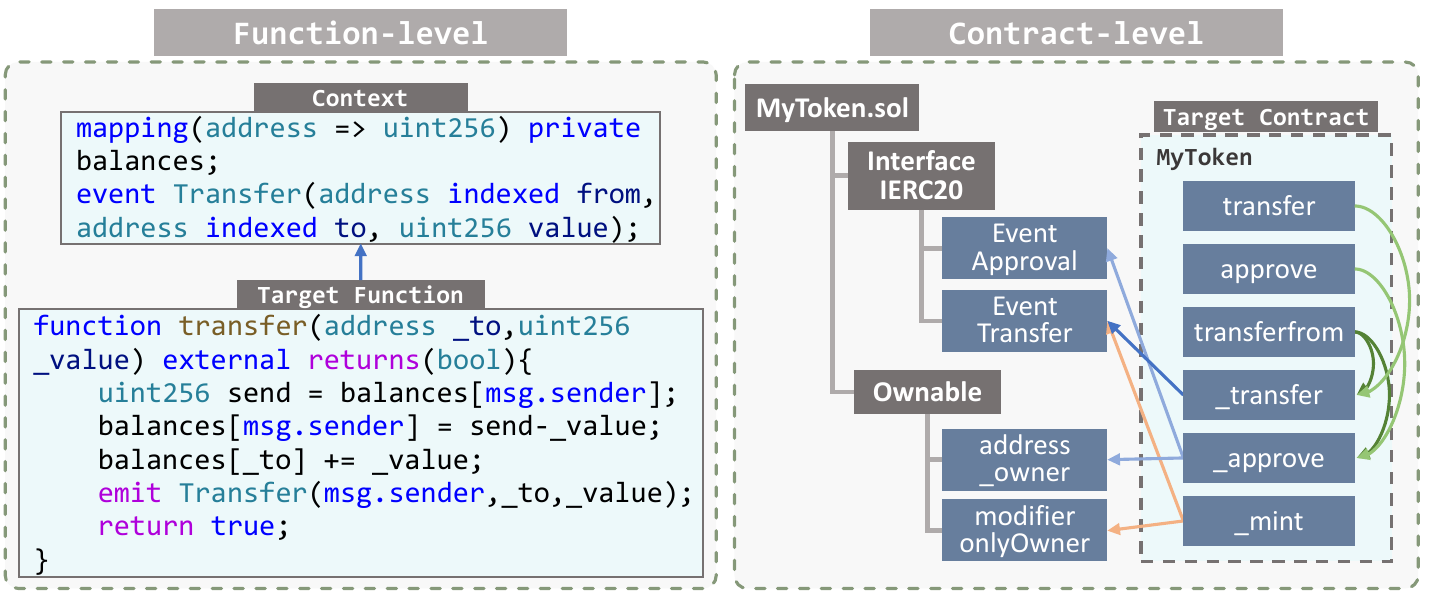}
    \caption{Function-Level vs. Contract-Level}
    \label{fig:compare}
\end{figure}

Notably, most existing research~\cite{peng2025soleval,chen2025solbench} focuses on function-level tasks that mainly assess syntactic correctness and limited contextual reasoning. 
As illustrated in Figure~\ref{fig:compare}, function-level tasks involve limited context compared to contract-level ones. For example, the target function \textit{transfer} depends only on relevant variables and events within the same contract. This limited scope means function-level generation does not reflect the full complexity of real-world development. In contrast, contract-level generation must handle the full contract, as in the case of \textit{MyToken}, which includes complex internal calls and dependencies on external contracts such as \textit{IERC20} and \textit{Ownable}. These factors substantially increase generation difficulty, highlighting the limitations of function-level benchmarks for practical evaluation.
Furthermore, current evaluation methods largely depend on synthetic test cases or fuzz testing techniques~\cite{peng2025soleval, daspe2024benchmarking}, which struggle to capture the dynamic constraints and behavioral patterns of real on-chain transactions. As a result, existing approaches suffer from notable limitations in both task granularity and evaluation methods, making them inadequate for providing a comprehensive analysis of LLM generation quality.

To address the gap in smart contract generation research, this study explores the more challenging and practical contract-level code generation task. Specifically, we focus on generating a single complete and functional contract selected from multi-contract Solidity files, based on a corresponding natural language description. To support this task, we manually construct \textit{\textbf{SolContractEval}}, the first contract-level code generation benchmark.
\textit{SolContractEval} comprises 124 generation tasks, with contract code sourced from Etherscan~\cite{etherscan2025}, covering nine mainstream application domains, e.g., decentralized finance (DeFi)~\cite{jensen2021introduction}, decentralized gaming (GameFi)~\cite{proelss2023gamefi}, and decentralized autonomous organizations (DAOs)~\cite{santana2022blockchain}. Each task input includes \textit{contract context dependencies}, a \textit{target contract framework}, and a clear \textit{task prompt}. To ensure data quality, each task input is independently annotated and cross-validated by two developers with three years of experience in smart contract development. Furthermore, to accurately and automatically assess the functional correctness of generated contracts, we introduce a \textbf{transaction replay mechanism} using real on-chain transactions. 
By comparing execution results and compilation success, this approach enables a practical evaluation of generation quality.

Building upon the proposed benchmark, we conduct the \textbf{\textit{first systematic evaluation}} of LLMs on contract-level code generation. Specifically, our experiments involve six widely used LLMs that vary in model sizes, underlying architectures, and source providers. For each generated code snippet, we integrate it back into the original contract file and assess the generation performance with two metrics: \textit{Compile@k}, which measures compilation correctness, and \textit{Pass@k}, which evaluates functional correctness.

Based on our empirical study, we summarize the following key findings: \textit{First}, Claude-3.7-Sonnet~\cite{anthropic2025claude} and o4-mini~\cite{openai2025o4-minis} demonstrate the strongest performance in contract-level code generation, followed by GPT-4o~\cite{openai2024GPT-4o}, DeepSeek-R1~\cite{guo2025deepseek}, and Qwen2.5-Coder-32b~\cite{hui2024qwen2}. \textit{Second}, current LLM shows weaker performance on Solidity contract-level generation than on class-level generation in general-purpose languages. \textit{Third}, models perform relatively well on the most common ERC20 contract type, but show significant shortcomings in generating more complex contracts like ERC721 and DeFi. \textit{Finally}, our error analysis reveals persistent limitations in handling Solidity-specific language features (e.g., compiler version compatibility, type system) and contextual dependency resolution across tested models.

In summary, this paper makes the following contribution:
\begin{itemize}
    \item The first benchmark \textit{SolContractEval} for contract-level code generation in Solidity, constructed from real-world on-chain contracts and enhanced with historical transaction replay for realistic and reliable functional validation
    \item The first systematic evaluation of mainstream LLMs on contract-level Solidity code generation, addressing a critical research gap in smart contract development.
    \item In-depth analysis of model-generated errors, revealing common failure patterns and providing implications for improving LLMs on Solidity-specific generation tasks.
    \item All code and data for this study are publicly available at https://github.com/ZJU-CTAG/SolContractEval.
\end{itemize}

\section{Background}

\subsection{Explanations of Terminologies}

\textbf{\textcolor{black}{$\blacksquare$} Solidity Source File.} Solidity source files follow a standardized modular architecture~\cite{contract2025}, which includes the following components: \textit{license identifier} (clarify intellectual property terms), \textit{pragma directive} (ensure version compatibility), \textit{import statements} (enable code modularization and reuse), and \textit{contract definitions} (the core entity for business logic).

\textbf{\textcolor{black}{$\blacksquare$} Transaction Replay.} Transaction replay verifies state changes by re-executing historical transactions~\cite{huang2024ethereum}. It restores account states to their pre-execution condition and re-runs the transaction in the same context. Since blockchain nodes usually retain only the latest world state, replaying transactions requires external sources such as archive nodes or state snapshots. In our study, we use this technique to extract post-execution state changes from the first 1,000 transactions for comparison and analysis.

\textbf{\textcolor{black}{$\blacksquare$} Events.} In Solidity, an event is a logging mechanism used to record critical operations and state changes~\cite{li2023understanding}. An event consists of a declaration and its parameters, and is triggered using the \textit{emit} keyword. Once emitted, the event data is written to the transaction receipt rather than the contract state, enabling low-cost persistence. Their immutable on-chain storage enables reliable asynchronous communication between smart contracts and off-chain systems.

\textbf{\textcolor{black}{$\blacksquare$} Solidity Compiler Versions.} As of May 2025, the Solidity compiler has officially released 8 major versions and 112 minor versions~\cite{solidity2025}. Mainstream smart contract development is currently concentrated between versions 0.4 and 0.8, with 0.4 (27 minors) and 0.8 (30 minors) being most prevalent.

\textbf{\textcolor{black}{$\blacksquare$} NatSpec.} Solidity supports \textbf{Ethereum Natural Specification Format (NatSpec)}, a specialized syntax for structured documentation of functions, parameters, and more~\cite{NatSpecFormatSolidity}.

\subsection{Distinctive Language Features of Solidity}

While Solidity borrows syntax from JavaScript and C++, it differs fundamentally from general-purpose languages in its technical design~\cite{wang2021empirical}, posing unique challenges for LLMs that perform well in conventional languages but struggle with Solidity’s specialized constraints.

\subsubsection{Gas Mechanism} Developers must explicitly manage computational resources, as each state-changing operation incurs a Gas cost. Unlike general-purpose languages optimized for CPU or memory, Solidity prioritizes Gas efficiency.

\subsubsection{Immutability} Solidity’s immutability requires rigorous verification before deployment, since any logic change demands redeploying the entire contract. This contrasts with hot-update mechanisms in traditional software.

\subsubsection{State Layering Mechanism} Solidity enforces a distinction between state storage and temporary memory: storage variables persist on-chain, occupying storage and incurring Gas costs, while memory variables exist only during transaction execution and are charged based on usage.

\subsection{Large Language Models for Code Generation}
Code generation, or natural language to code (NL2Code), aims to generate executable code from natural language descriptions automatically and has attracted growing attention from both academia and industry~\cite{jiang2024survey, zan2022large, chen2024b4}. Large language models (LLMs), including general-purpose models such as GPT-4~\cite{openai2024GPT-4o} and Claude~\cite{anthropic2025claude}, as well as code-focused models like StarCoder 2~\cite{lozhkov2024starcoder} and Code Llama~\cite{roziere2023code}, have shown strong performance on code generation benchmarks such as HumanEval~\cite{chen2021evaluating,chen2024jumpcoder}. This success largely stems from pretraining on large corpora of high-quality code, enabling models to learn general syntax and semantics across languages. 

However, training data for these models is heavily imbalanced. Popular languages like Python and JavaScript are richly represented. In contrast, Solidity, the primary language for smart contracts, comprises just 0.01\% of Stack v2, the pretraining dataset used by models such as StarCoder 2~\cite{lozhkov2024starcoder}. This limited exposure raises concerns about whether current models can generalize effectively to Solidity code generation.

\section{Benchmark Construction}

\subsection{Overview of SolContractEval}

\textbf{SolContractEval} is a benchmark dataset for contract-level Solidity code generation, comprising 124 generation tasks. Specifically, \textbf{contract-level code generation} means generating a complete, functional target contract from a multi-contract Solidity file based on a natural language description. In our setting, the contract to be generated is the \textbf{target contract}, and the original Solidity file it’s drawn from is the \textbf{original contract}. All original contracts are real-world contracts deployed on the Ethereum mainnet~\cite{EthereumorgCompleteGuidea}, covering major domains like Decentralized Finance (DeFi), Non-Fungible Tokens (NFTs), and GameFi. The table~\ref{tab:statistics} presents the overall statistics of our benchmark, as well as detailed statistics for the three contract categories with the highest proportions within the benchmark.

\begin{table}[htbp]
    \centering
    \caption{Statistics of SolContractEval Benchmark.\\SCs:Smart Contracts. Txs:Transactions per Test.}
    \label{tab:statistics}
    \begin{adjustbox}{width=\columnwidth}
    \begin{tabular}{lcccccc}
    \hline
        ~ & \#Tasks & \makecell{\#Funcs} & \makecell{\#Tokens} & \makecell{\#Context\\SCs} & \makecell{\#Context\\Tokens} & \makecell{\#Txs}  \\ \hline
        ERC20 & 52 & 11.23  & 1206.79  & 4.33  & 1074.04  & 1000  \\
        DeFi & 28 & 9.75  & 1117.82  & 4.75  & 702.50  & 1000  \\ 
        ERC721 & 23 & 17.26  & 1628.35  & 16.22  & 4841.30  & 1000  \\ 
        Others & 21 & 8.00  & 1124.38  & 6.19  & 1549.33  & 1000  \\ \hline
        \textbf{Overall} & 124 & 11.47  & 1250.93  & 6.94  & 1769.40  & 1000  \\ \hline
    \end{tabular}
    \end{adjustbox}
\end{table}

Each task consists of three components: (1) \textbf{input information} that guides the language model to generate the target contract; (2) a \textbf{test suite}, which verifies the functional correctness of the generated code; and (3) the corresponding \textbf{original contract}, which serves as a reference for both functionality and structural design.

\begin{figure}[!htbp]
    \centering
    \includegraphics[width=\columnwidth, clip]{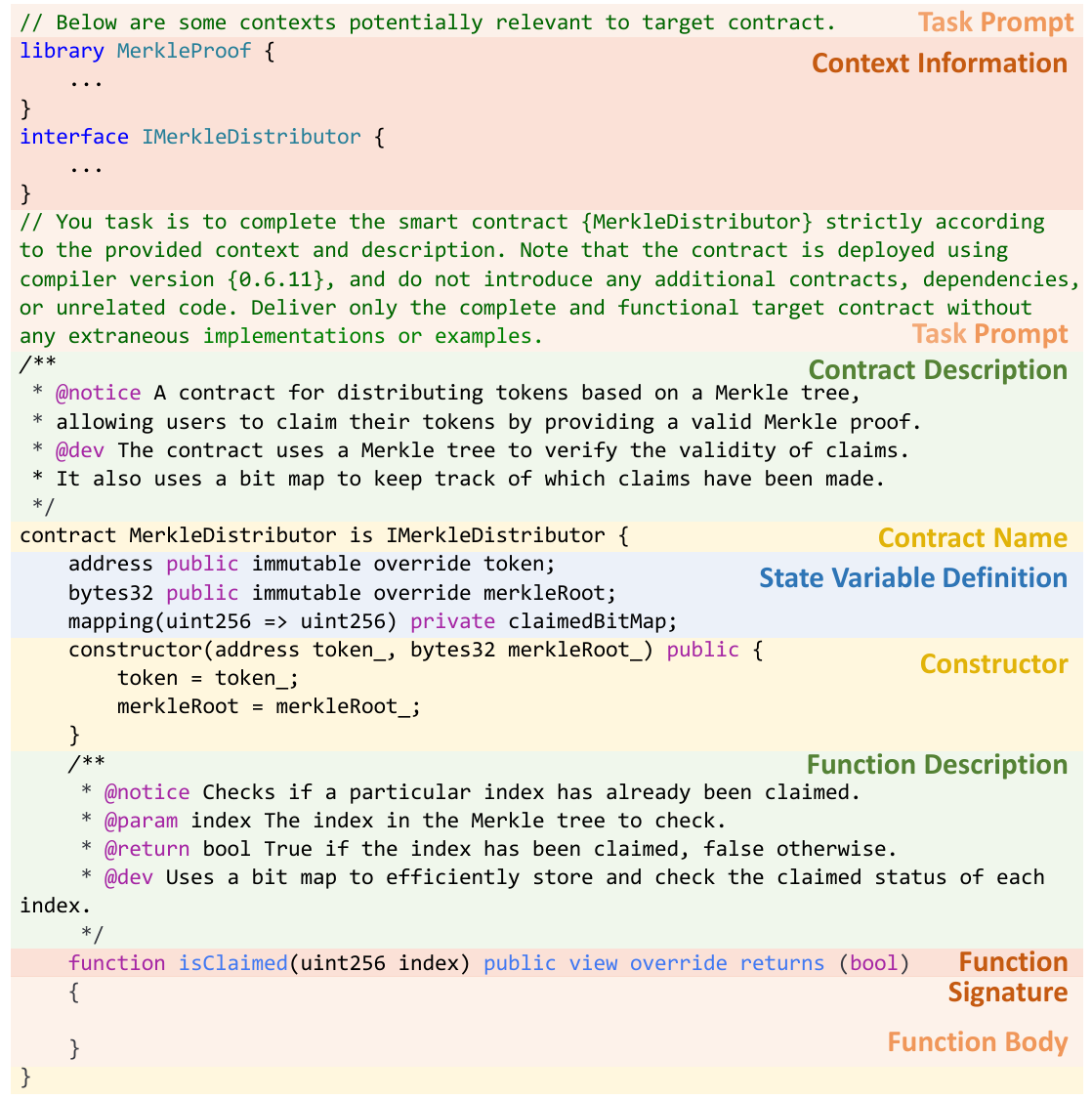}
    \caption{An Example of the Input Information}
    \label{fig:contract_framework}
\end{figure}

To guide the large language model in generating realistic and functional contract code, each code generation task takes the following three components as input, as illustrated in Figure~\ref{fig:contract_framework}:
(1) \textbf{context information}, which includes relevant functions, variables, and contract definitions extracted from real-world repositories;
(2) \textbf{target contract framework}, a structured scaffold that outlines  contract-level elements;
(3) \textbf{task prompt}, which specifies the task requirements, including the desired contract name and the compiler version for deployment.
The target contract framework serves as the structural backbone for code generation. It contains the contract name, state variable definitions, constructor, function signatures, and their corresponding descriptions. Since this framework is directly used to replay historical transactions and run automated tests, the correctness of generated code critically depends on maintaining consistent interfaces. To enforce this, we preserve the original function signatures and only leave the function bodies blank for the large language model to complete, while keeping all other structural elements intact.

We feed the pre-constructed input information into the model to obtain its response. From the model's output, we extract the corresponding target contract implementation and replace the original implementation within the contract structure to form a new generated contract. Both the original contract and the generated contract are then evaluated using the test suite. The evaluation process first checks compilation success, followed by a comparison of execution states, event emissions, and storage results, to further assess the functional correctness and behavioral consistency of the generated contract.

\subsection{Benchmark Construction}

\begin{figure*}[!tbp]
    \centering
    \includegraphics[width=\textwidth]{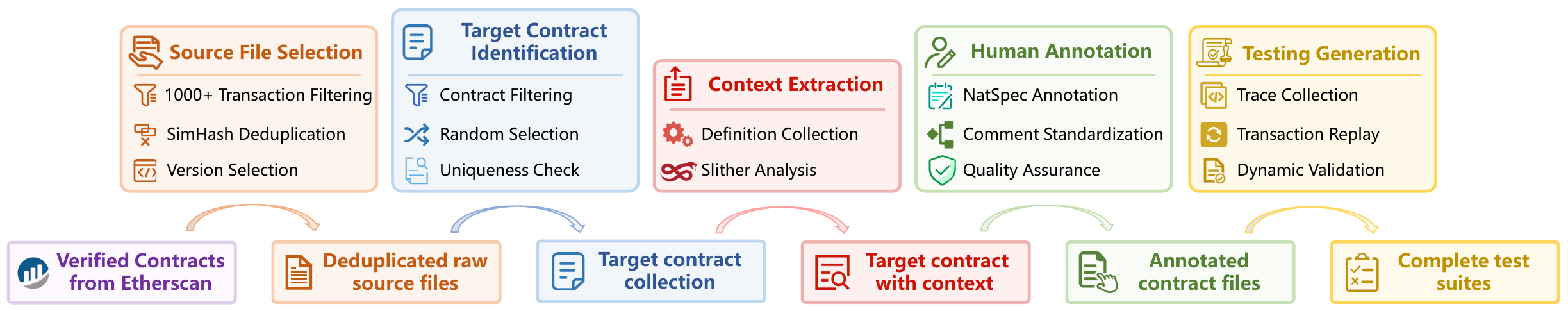}
    \caption{Overview of SolContractEval Construction Process}
    \label{fig:construction}
\end{figure*}

This section details the construction of our benchmark for evaluating Solidity code generation, as illustrated in Figure~\ref{fig:construction}, covering source file selection, target contract identification, context extraction, human annotation, and testing generation.

\subsubsection{\textbf{Source File Selection}}
We begin by collecting verified smart contract source files from the dataset used in the study by Ajibola et al.~\cite{ajibola2025smartpattern}, sourced from Etherscan~\cite{etherscan2025}, the largest repository of open-source, verified contract implementations. To ensure active usage and suitability for transaction-based tests, we filter contracts with more than 1,000 historical transactions, as Zhang et al.\cite{zhang2025automated} demonstrated that this threshold covers common contract functionalities sufficiently. Only contracts with publicly available and compiler-verified source code are retained to enable precise static and dynamic analysis.
This process yields 4,640 contract source files, which are then deduplicated using a SimHash~\cite{SimHash2024} algorithm to eliminate syntactically near-identical implementations (e.g., forks or minimal modifications), resulting in 1,278 unique contract source files. 
To ensure balanced representation across compiler versions, we group deduplicated contracts by their declared Solidity version and randomly select up to 50 source files per group. If a group has fewer than 50, all files are included. This yields 50 files each for the commonly used 0.4.x and 0.8.x versions, and 18, 20, and 16 files for versions 0.5.x, 0.6.x, and 0.7.x, respectively.

\subsubsection{\textbf{Target Contract Identification}}
From each collected Solidity source file, we select one \textbf{target contract} for the code generation task, mirroring real-world development practices where developers typically implement individual contracts sequentially.
To ensure the quality and reliability of the benchmark evaluation, we exclude interface contracts, which contain only function declarations without implementations, abstract contracts, which rely on child contracts to complete their logic, and contracts that only define constructors without implementing any additional functional logic.
After filtering, we randomly select one contract from each source file as the target for generation. Random selection minimizes human bias and enhances the benchmark's objectivity. Finally, we performed deduplication to ensure each contract in the final dataset was unique, preventing potential statistical bias during evaluation, ultimately obtaining 124 distinct contracts.

\subsubsection{\textbf{Context Extraction}}

While our task focuses on contract-level code generation, maintaining source file context is crucial for compilation and execution. Missing or unresolved dependencies, such as libraries, inherited contracts, or utility functions, can cause compilation failures or runtime errors due to undefined references. As illustrated in the contract-level dependency diagram in Figure~\ref{fig:compare}, the arrows represent dependency relationships. For example, the target contract \textit{MyToken} depends on external interfaces like \textit{IERC20} and \textit{Ownable}; if these are not provided, the model may generate undefined references, resulting in compilation errors.

\begin{algorithm}[!htbp]
    \caption{Algorithm to Extract Context for Contract}
    \label{alg:context}
    \renewcommand{\algorithmicrequire}{\textbf{Input:}}
    \renewcommand{\algorithmicensure}{\textbf{Output:}}
    
    \begin{algorithmic}[1]
        \REQUIRE $SC$, $TargetContract$ \COMMENT{Input: Smart contract and target contract block}
        \ENSURE $Context$ \COMMENT{Output: Context of target contract}

        \STATE $FCG \leftarrow \text{BuildFCG}(SC)$ 
        \STATE $Funcs \leftarrow \text{GetFuncs}(TargetContract)$  
        \STATE $ExtFuncs \leftarrow \emptyset$  
        \STATE $Context \leftarrow \emptyset$  

        \FOR{each function $f \in Funcs$}
            \STATE $Deps \leftarrow \text{GetExternalDeps}(f)$  
            \FOR{each dependency $dep \in Deps$}
                \IF{$dep$ not in $ExtFuncs$}
                    \STATE $ExtFuncs \leftarrow ExtFuncs \cup \{dep\}$  
                \ENDIF
            \ENDFOR
        \ENDFOR

        \FOR{each function $f \in ExtFuncs$}
            \STATE $DepContract \leftarrow \text{FindContract}(f)$  
            \STATE $Range \leftarrow \text{GetCodeRange}(DepContract, f)$ 
            \STATE \text{Add}($f$, $Range$) \text{ to } $Context$  
        \ENDFOR

        \STATE \text{Return } $Context$  
    \end{algorithmic}
\end{algorithm}

Algorithm~\ref{alg:context} outlines the systematic process for context extraction, ensuring that generated contracts remain compilable, executable, and functionally equivalent to their original counterparts. The process begins by parsing the target Solidity source file to collect all relevant definitions, including imports, global variables, functions, and inheritance hierarchies. This step aims to capture every element necessary for the contract to operate as intended within its original environment.

Next, we employ Slither~\cite{feist2019slither}, a static analysis framework for Solidity, to identify external references that the target contract depends on but does not define internally. These dependencies may include externally invoked functions, inherited contracts, or library methods referenced but not implemented within the target contract.

For each external dependency detected, we locate its definition within the Solidity file’s contracts or libraries and extract the minimal set of required components, specifically the functions, variables, or structures used by the target contract. We isolate the code ranges of referenced functions to avoid including extraneous logic. By incorporating only essential dependencies, we ensure correctness while reducing overhead.

\subsubsection{\textbf{Human Annotation}}
In Solidity code generation tasks, the quality of the prompt significantly influences the performance of the generated outputs. To ensure accurate and standardized function descriptions, we engage two developers with at least three years of Solidity development experience to manually write the descriptions for all functions and target contracts in each code generation task. The inclusion of human-annotated comments serves two purposes: 1) to minimize the potential memorization effects of LLMs, as they may have encountered the original comments during pre-training. 2) to provide accurate and consistent function descriptions, ensuring high-quality annotations for the benchmark.

The annotation process follows key standards. First, all comments must comply with the NatSpec documentation format officially recommended by Solidity, using a structured style. Second, to minimize noise, non-functional comments such as copyright notices or irrelevant content are excluded. Third, each function's purpose and parameters must be described with precision and completeness.
Each task instance undergoes dual independent annotation based on these criteria, followed by cross-verification. In cases of disagreement, annotators resolve conflicts through discussion to reach consensus.

\subsubsection{\textbf{Testing Generation}}

Evaluating the correctness of smart contract generation is challenging due to the dynamic and stateful nature of contract execution. Unlike traditional code, smart contracts are tightly coupled with the blockchain, where correctness depends not only on isolated function behavior but also on interactions with historical on-chain data, external calls, and mutable state. This complexity renders static analysis and manually crafted test cases insufficient, as they fail to capture real-world execution scenarios.

\begin{figure}[htbp]
    \centering
    \includegraphics[width=\columnwidth, clip]{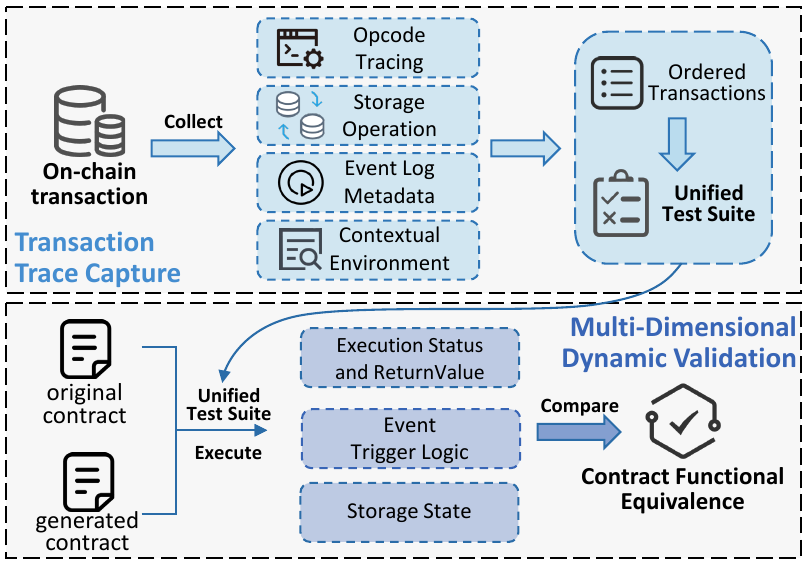}
    \caption{The Technical Workflow of Testing Generation}
    \label{fig:architecture}
\end{figure}

To address these challenges, we introduce a transaction-driven dynamic validation framework inspired by Zhang et al.\cite{zhang2025automated}. The key idea is to leverage real transaction data to construct realistic test suites that mirror actual contract behavior in production. We implement this by replaying authentic transaction sequences on generated contracts and dynamically monitoring their execution to assess correctness. As shown in Figure~\ref{fig:architecture}, this approach enables robust, automated validation under realistic conditions and improves evaluation reliability.

\paragraph{\textbf{On-Chain Transaction Trace Capture}}
We collect the target contract's historical transaction data through an Ethereum archive node, capturing the complete lifecycle of its first 1,000 transactions. For each transaction, we extract four key categories of execution information to support fine-grained behavioral analysis:
\begin{itemize}
    \item \textbf{Opcode Tracing:} The \textit{debug\_traceTransaction} interface is used to record the sequence of EVM instructions.
    \item \textbf{Storage Operation Snapshot:} All persistent storage \textit{read/write} operations for each transaction are recorded in a structured format as \textit{⟨slot, oldValue, newValue⟩} triplets.
    \item \textbf{Event Log Metadata:} Event signatures, indexed parameters, and non-indexed parameters are parsed according to the Application Binary Interface (ABI) specification.
    \item \textbf{Contextual Environment:} The system records the full execution context, including the sender address (\textit{msg.sender}), and block-level context parameters (\textit{block.timestamp, block.number}).
\end{itemize}

To ensure high-quality validation, we integrate the first 1,000 historical transactions into a single unified test sequence, strictly preserving their original on-chain execution order. We avoid splitting these transactions into multiple isolated test cases, as this would break critical dependencies between transactions. Treating them as a coherent whole allows us to more comprehensively capture the behavioral context of the target contract, ensuring the test sequence can reliably assess the functional correctness of the generated contract in realistic environments. Empirical evidence from Zhang et al.~\cite{zhang2025automated} further supports this approach, showing that the first 1,000 transactions typically cover commonly used functionalities of a contract, validating the effectiveness of our testing design.

We implement this on the Hardhat~\cite{HardhatEthereumDevelopment} local testing network, resetting blockchain state before execution to avoid cross-test interference. Each transaction is replayed with original contextual parameters (e.g., \textit{msg.sender}, \textit{msg.value}), ensuring consistent state evolution and reliable functional validation.

\paragraph{\textbf{Multi-Dimensional Dynamic Validation}}
We employ a unified test suite to verify functional equivalence between original and generated contracts across three systematic dimensions. Functional correctness is confirmed only when all three dimensions exhibit consistent behavior.

\begin{itemize}
    \item \textbf{Execution Status and Output Consistency:} The system compares the status field in transaction receipts, requiring strict correspondence between success and revert outcomes. Additionally, the return values are validated byte-by-byte to ensure semantic equivalence of outputs.
    \item \textbf{Event Trigger Logic Equivalence:} Based on the contract ABI, event structures are parsed and validated through a dual-constraint mechanism. First, generated contract event signatures must precisely match the original. Next, parameters undergo rigorous comparison, including hash verification of indexed parameters and value-by-value checks for non-indexed ones. These checks ensure structural and content-level accuracy in event behavior.
    \item \textbf{Storage State Consistency:} Key state variables (e.g., \textit{totalSupply}, \textit{allowance}) are hashed before and after each transaction using SHA3-256. A state evolution chain is constructed to verify consistent state transitions across the original and generated contracts.
\end{itemize}

\section{Experimental Setup}

Using SolContractEval, we conduct a comprehensive study to evaluate LLMs on contract-level Solidity code generation, guided by the following research questions.

\begin{itemize}
    \item \textbf{RQ1 (Overall Correctness):} How do LLMs perform on Solidity contract-level code generation?
    \item \textbf{RQ2 (Performance across Different Contract Categories):} How does model performance vary across smart contract categories?
    \item \textbf{RQ3 (Incorrect Case Analysis):} What are the common errors during Solidity contract-level code generation?
\end{itemize}

\subsection{Studied LLM}
In this study, we evaluate six state-of-the-art LLMs that have demonstrated strong performance in code generation and are widely adopted in recent research~\cite{quan2025codeelo, li2024can}.
For closed-source models, we select leading commercial options, including OpenAI’s GPT-4o~\cite{openai2024GPT-4o} and o4-mini~\cite{openai2025o4-minis}, Anthropic’s Claude-3.7-sonnet~\cite{anthropic2025claude}, and Google DeepMind’s Gemini-2.0-flash-exp~\cite{Google2024gemini2}.
For open-source models, we include DeepSeek-AI’s DeepSeek-R1~\cite{guo2025deepseek} and Alibaba’s Qwen2.5-Coder-32b-instruct~\cite{hui2024qwen2}.
These models have exhibited strong performance across multiple code generation benchmarks, demonstrating their capabilities in practical applications.

\subsection{Evalution Metrics}
\label{setup:metrics}
In this study, we adopt the \textit{Pass@k} and \textit{Compile@k} metrics to evaluate the generated Solidity code in terms of functional and compilation correctness, respectively. Both are based on the widely used \textit{Pass@k} framework~\cite{chen2021evaluating, du2023classeval, yu2024codereval}. For each task, the model generates $n$ candidate code samples, and the task is considered successful if at least one of the top-$k$ samples meets the evaluation criterion, passing all unit tests for \textit{Pass@k} or compiling successfully for \textit{Compile@k}. To mitigate the variance introduced by random sampling, we adopt the unbiased estimator from the HumanEval benchmark~\cite{chen2021evaluating}:

\begin{equation}
    Compile@k/Pass@k=\underset{\text { Problems }}{\mathbb{E}}\left[1-\frac{\binom{n-c}{k}}{\binom{n}{k}}\right]
    \label{eq1}
\end{equation}
where $n$ is the total number of generated samples, $c$ is the number of correct (i.e., passing or compilable) samples, and $k$ is the sampling budget per task.

To compare contract-level generation with function-level generation, we introduce two evaluation metrics for this setting: \textit{Function Pass Rate (FPR)} and \textit{Contract Full Pass Rate (CFPR)}. Each has two variants, with subscripts ``c'' and ``p'' indicating compilation and test success, respectively. 

Function Pass Rate (FPR) measures the accuracy of the model when generating individual functions. Specifically, FPR is the ratio of successfully compiled (or tested) functions $P$ to the total number of generated functions $M$, i.e., $FPR_c/FPR_p = \frac{P}{M}$. This metric reflects function-level Pass@1, assuming one generation attempt per function.

Contract Full Pass Rate (CFPR) measures whether a contract can compile and pass all tests when each function is generated independently. If all individually generated functions can be re-integrated into the original contract and still compile (or pass tests), the contract is considered “fully passed”. Thus, CFPR is the ratio of fully passed contracts $K$ to the total number of contracts $N$, i.e., $CFPR_c/CFPR_p = \frac{K}{N}$.

\subsection{Implementation Details}
We evaluate the generated contracts using two primary metrics: Compile@k, which measures compilation correctness, and Pass@k, which assesses functional correctness. Following recent practice~\cite{du2023classeval}, we adopt two sampling strategies for code generation: (1) \textbf{nucleus sampling}, in which five candidate code completions are generated per task using a temperature of 0.2 and top\_p of 0.95\cite{chen2021evaluating}; and (2) \textbf{greedy sampling}, in which a single deterministic completion is generated using a temperature of 0. For both metrics, greedy sampling is used when $k = 1$, while nucleus sampling is applied when $k > 1$.
To ensure fair model comparison, we invoke APIs with consistent hyperparameter settings. All experiments are run on a high-performance server with Ubuntu 20.04.5 LTS, 128 Intel(R) Xeon(R) Platinum 8358P CPUs @ 2.60GHz, and 8 NVIDIA A800 GPUs (80GB each), providing a stable environment for large-scale code generation and evaluation.

\section{Results}

\subsection{RQ1: Overall Correctness}
\label{result:overall}
\begin{table*}[t]
    \centering
    \caption{Overall Results on SolContractBench.}
    \label{tab:overall}
    \begin{tabular*}{0.9\linewidth}{l c c c  c c c }
    \hline
        \textbf{Model} & \textbf{Compile@1(\%)} & \textbf{Compile@3(\%)} & \textbf{Compile@5(\%)} & \textbf{Pass@1(\%)} & \textbf{Pass@3(\%)} & \textbf{Pass@5(\%)} \\ \hline
        gemini-2.0-flash-exp & 63.55 & 69.66 & 72.03 & 29.03 & 32.66 & 33.87 \\
        qwen2.5-coder-32b-instruct & 68.54 & 78.62 & 84.67 & 25.97 & 33.63 & 35.48 \\ 
        deepseek-R1 & 67.74 & 81.77 & 86.29 & 29.84 & 37.10 & 40.32 \\ 
        gpt-4o & 68.54 & 90.32 & 94.35 & 33.39 & 42.10 & 45.16 \\ 
        o4-mini & 81.45 & \textbf{93.54} & 95.97 & 35.00 & 42.18 & 44.35 \\ 
        claude-3-7-sonnet-20250219 & \textbf{85.48} & \textbf{93.54} & \textbf{96.77} & \textbf{40.65} & \textbf{46.05} & \textbf{49.19} \\ \hline
    \end{tabular*}
\end{table*}

Our comprehensive evaluation of six state-of-the-art LLMs on the SolClassEval benchmark reveals their comparative performance, as detailed in Table~\ref{tab:overall}.

\textbf{Compilation Correctness.} Claude-3-7-Sonnet performs best, leading in both \textit{Compile@1} (greedy sampling) and \textit{Compile@k} (nucleus sampling) at the function level. o4-mini follows closely, with only a 4.03\% lower \textit{Compile@1} but nearly matching Claude-3-7-Sonnet at \textit{Compile@k}, indicating comparable robustness. GPT-4o, DeepSeek-R1, and Qwen2.5-Coder-32b come next, all maintaining \textit{Compile@1} rates around 68\%. Notably, GPT-4o shows the greatest gain with nucleus sampling, suggesting strong performance under diverse decoding strategies. In contrast, Gemini-2.0-Flash-Exp performs worst due to recitation errors (see Section~\ref{discussion:recitation}), which frequently lead to compilation failures under both sampling methods.

\textbf{Functional Correctness.} In terms of functional correctness, Claude-3-7-Sonnet again leads, with a \textit{Pass@1} score 5.65\% higher than the second-best model. Although o4-mini ranks highly in compilation success, its functional correctness is only slightly better than GPT-4o, indicating room for improvement in logical accuracy and context comprehension. Deepseek-R1, Qwen2.5-Coder-32b, and Gemini-2.0-Flash-Exp perform relatively poorly in functional implementation, with Qwen2.5-Coder-32b ranking last in \textit{Pass@1}. Interestingly, Gemini-2.0-Flash-Exp, affected by the "recitation" issue, shows comparable functional correctness to Qwen2.5-Coder-32b.

\begin{center}
    \begin{myboxc} \textbf{Findings 1:} In Solidity contract-level code generation, Claude-3-7-Sonnet and o4-mini lead, followed by GPT-4o, DeepSeek-R1, and Qwen2.5-Coder-32b, with Gemini-2.0-Flash-Exp performing the worst.
    \end{myboxc}
\end{center}

\textbf{Contract-level code generation v.s. Function-level code generation.} To assess the greater challenge posed by contract-level generation compared to function-level generation, we conduct function-level experiments on the benchmark using all six models. Specifically, for each target contract, we treat each function as an independent generation target, masking only the corresponding function body while preserving the implementation of other functions. This ensures that the contextual information remains consistent with the contract-level setting. 

\begin{table}[htbp]
    \centering
    \caption{Performance Comparison: Function-level vs. Contract-level.}
    \label{tab:function}
    \begin{adjustbox}{width=1\columnwidth}
    \begin{tabular}{l| c c c| c c c}
        \hline
        \multirow{2}{*}{\textbf{Model}} & \multicolumn{2}{c}{Function} & Contract & \multicolumn{2}{c}{Function} & Contract\\
        & FPR$_c$ & CFPR$_c$ & Compile@1 & FPR$_p$ & CFPR$_p$ & Pass@1\\[2pt] \hline
        gemini & 82.77 & 68.18 & 63.55 & 68.67 & 55.45 & 29.03 \\
        qwen2.5 & 92.12 & 76.14 & 68.54 & 71.15 & 54.12 & 25.97 \\
        deepseek & 92.62 & 77.53 & 67.74 & 67.09 & 53.37 & 29.84 \\
        gpt-4o & 93.22  & 85.36  & 68.54  & 72.28  & 60.16  & 33.39   \\ 
        o4-mini & 95.08  & 84.23  & 81.45  & 71.89  & 61.73  & 35.00   \\ 
        claude & 96.23  & 86.71  & 85.48  & 80.42  & 65.85  & 40.65   \\ \hline
    \end{tabular}
    \end{adjustbox}    
\end{table}

As shown in Table~\ref{tab:function}, all models achieve significantly better performance under the function-level setting across various evaluation metrics. For example, the $FPR_p$, which reflects model accuracy on individual function generation, is higher than \textit{Pass@1} by 38.52\% on average. Claude-3-7-Sonnet achieves $FPR_c$/$FPR_p$ scores of 96.23\%/80.42\%, while its \textit{Compile@1/Pass@1} scores under the contract-level setting are only 85.48\%/40.65\%, indicating that models tend to perform better on isolated function tasks, possibly due to the more clearly defined context.

Furthermore, we compare the overall contract correctness under both settings by focusing on the CFPR for function-level and the corresponding metrics for contract-level. CFPR measures whether all functions within a contract can be independently compiled or pass their corresponding tests. The gap between the two settings is substantial. For example, GPT-4o achieves $CFPR_c$/$CFPR_p$ scores of 85.36\%/60.16\% at the function level, compared to just 68.54\%/33.39\% (\textit{Compile@1}/\textit{Pass@1}) at the contract level. This suggests that while models are capable of generating individual functions well, generating multiple functions simultaneously often leads to semantic inconsistencies or interaction errors, resulting in a significantly higher failure rate. 

\begin{center}
    \begin{myboxc} \textbf{Findings 2:} Higher function-level scores show isolated tasks are easier, while the gap with contract-level results highlights the challenge of maintaining inter-function consistency, essential in real-world development, making contract-level generation a more realistic and challenging setting.
    \end{myboxc}
\end{center}

\textbf{Compare to Other Benchmark.} 
\label{result:compare}
To compare Solidity benchmarks with those for general-purpose programming languages, we select ClassEval~\cite{du2023classeval}, a class-level benchmark of similar granularity to ours. We evaluate the GPT-3.5~\cite{IntroducingChatGPT2024} model used in the ClassEval paper on SolContractEval to ensure fair comparison. The results show a performance gap: GPT-3.5 achieves 29.6\%/34.9\%/36.0\% on ClassEval's \textit{Pass@1/3/5} for class-level generation, while its performance on SolContractEval drops to 12.1\%/15.6\%/16.9\%, respectively. This substantial gap highlights the unique challenges of smart contract generation compared to general-purpose code and underscores the importance of SolContractEval in evaluating LLM capabilities for Solidity code generation.

\begin{center}
    \begin{myboxc} \textbf{Findings 3:} Compared to class-level code generation in general-purpose languages, current LLMs show weaker performance on Solidity contract-level generation, suggesting that contract-level tasks may pose additional challenges. This underscores the need for specialized benchmarks to better assess model capabilities in this domain.
    \end{myboxc}
\end{center}

\subsection{RQ2: Performance across Different Contract Categories}
\begin{figure}[htbp]
    \centering
    \includegraphics[width=\columnwidth, clip]{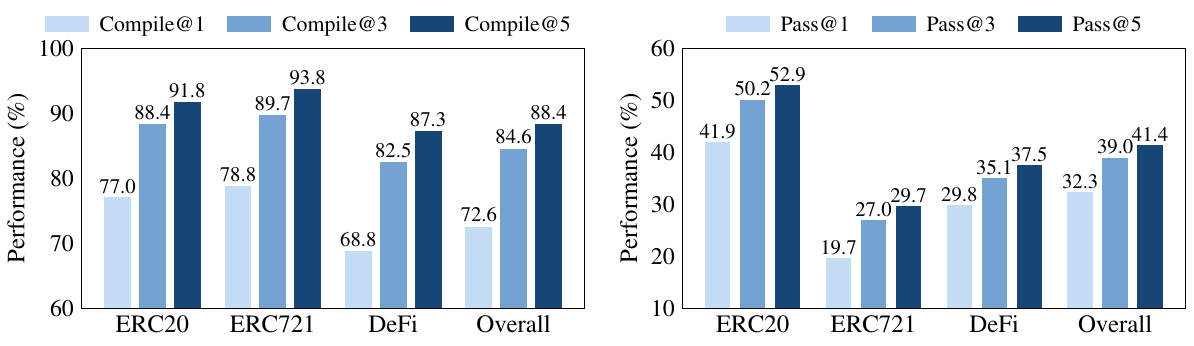}
    \caption{Model Performance Across Different Contract Categories}
    \label{fig:type}
\end{figure}

To analyze performance variations across contract categories, we calculate metrics for each category, select the three most prevalent by sample size, compute the model's average performance across them, and compare it with overall performance, as shown in Figure~\ref{fig:type}.

\textbf{ERC20.} In this benchmark test, the ERC20 category accounts for the largest proportion, reaching 41.9\%. The models exhibit better performance in this category compared to the overall average, with a slight improvement in compilation success rates and a substantial increase of 6.5\% in \textit{Pass@1}. This indicates that the models have a strong understanding of the ERC20 standard, which is widely adopted in practical development and features a stable, well-defined interface. This maturity enables the models to effectively learn diverse implementation patterns and best practices from training data. This extensive knowledge enhances the models' ability to generate targeted and accurate code, resulting in notable improvements in both compilation success rates and functional performance.

\textbf{ERC721.} Similarly, ERC721, another standardized token category, accounts for 18.6\% of the benchmark. Among the six models tested, four demonstrate significantly higher compilation success rates than the overall average, with overall performance surpassing that of ERC20. Only Qwen2.5-Coder-32b and Gemini-2.0-Flash-Exp show slightly inferior performance. However, in terms of functional correctness, ERC721 is the worst-performing category, with only Claude-3-7-sonnet reaching the overall average, while the remaining models exhibit varying degrees of decline.

\textit{Reason Analysis.} The performance difference between ERC20 and ERC721 contracts stems from several factors. ERC20 contracts cover multiple compiler versions (0.4 to 0.8), posing challenges for models to handle syntax variations across versions. In contrast, ERC721 contracts mainly target version 0.8, avoiding version compatibility issues and yielding relatively higher compilation success. However, ERC721 involves more complex logic and dependencies, such as ownership management, authorization controls, and security checks. These features are difficult to capture with simple patterns, making models prone to logical errors. Moreover, ERC721 typically depends on additional interfaces like ERC165 and IERC721Metadata, increasing generation complexity and negatively affecting functional correctness.

\textbf{DeFi.} In the benchmark, DeFi contracts account for a notable proportion (22.58\%). The models show a slight decline in compilation success rates for this category, with a 3.8\% decrease in the \textit{Compile@1} metric. This may be due to complex blockchain operations commonly used in DeFi contracts, such as \textit{delegatecall} and \textit{abi.encodePacked}, which can cause syntax or type errors during model generation, resulting in compilation failures. Nevertheless, regarding functional correctness, the models' performance on DeFi contracts is consistent with the overall level, indicating the models have acquired a reasonable understanding of core logic and common patterns in this category.

For other categories (such as Gaming and Proxy contracts), the number of samples is relatively small, each accounting for less than 10\%. As a result, the evaluation results in these categories may exhibit certain randomness.

\textit{Conclusion.} Standardized tokens like \textbf{ERC20} provide unified interfaces and mature implementations that help models generate code with high compilation success and functional correctness. In contrast, non-standard contracts (e.g., DeFi, Gaming) lack consistent patterns and diverse logic, making generation more error-prone and less reliable. This variability remains a key challenge for improving model output quality.

\begin{center}
    \begin{myboxc} \textbf{Findings 4:} Models excel in generating simple standardized contracts like ERC20, but struggle with complex logic in ERC721 and diverse operations in DeFi contracts, highlighting the challenges of non-standardized code.
    \end{myboxc}
\end{center}

\subsection{RQ3: Incorrect Case Analysis}
\label{result:case}
To help developers better leverage LLMs for smart contract code generation, we further analyze the common issues encountered in current models. Specifically, we examine code artifacts from the 124 generation tasks and categorize the identified issues into the following two types for detailed analysis: \textit{compilation errors} and \textit{testing errors}.

\begin{table}[!ht]
    \centering
    \caption{Syntax errors of LLM-generated code.}
    \label{tab:compile}
    \begin{tabular}{l r}
        \toprule
        \textbf{Category} & \textbf{Proportion} \\
        \midrule
        
        \textbf{ParserError} & \textbf{37.21\%}  \\ 
        \quad Event triggering method error & 22.48\%  \\ 
        \quad Missing parentheses/semicolon & 6.20\%  \\ 
        \quad Syntax structure error & 5.43\%  \\ 
        \quad Receive function version issue & 3.10\%  \\ 
        \midrule
        
        \textbf{TypeError} & \textbf{36.43\%}  \\ 
        \quad Override keyword conflict & 8.53\%  \\ 
        \quad Contract member unavailable & 6.20\%  \\ 
        \quad Type conversion error & 6.20\%  \\ 
        \quad Function parameter count mismatch & 5.43\%  \\ 
        \quad Function parameter type mismatch & 4.65\%  \\ 
        \quad Pure/View keyword conflict & 2.33\%  \\ 
        \quad Storage type conversion error & 1.55\%  \\ 
        \quad Member access error & 1.55\%  \\ 
        
        \midrule
        \textbf{DeclarationError} & \textbf{21.71\%}  \\ 
        \quad Undefined identifier & 15.50\%  \\ 
        \quad Variable redefinition & 4.65\%  \\ 
        \quad Call to invisible function & 1.55\%  \\ 
        
        \midrule
        
        \textbf{Other Errors} & \textbf{2.33\%}  \\ 
        \quad Stack depth limit & 2.33\%  \\ 
        \quad Incorrect association with function parameter & 1.55\%  \\ 
        \quad Others & 0.78\% \\
        \bottomrule
    \end{tabular}
\end{table}

\subsubsection{\textbf{Compilation Errors}} In smart contract development, compilation errors involve not only syntax issues but also potential deficiencies in handling specific risk scenarios. %
We systematically analyze compilation errors by first identifying the compiler version used in on-chain contracts and compiling corresponding generated contracts with that version, collecting 985 error logs. Using open card sorting\cite{lewis2010open}, we categorize these into 5 major types and 17 subcategories (Table~\ref{tab:compile}).
Among the major error types, \textit{ParserError} (37.21\%) and \textit{TypeError} (36.43\%) dominate, while \textit{DeclarationError} (21.71\%) is also frequent, together accounting for over 95\% of all errors. These results suggest that current models often fail at basic syntax understanding, type handling, and declaration correctness during contract-level code generation.
Further analysis of the subcategory errors reveals the following key issues:

\textbf{Inadequate Adaptation to Compiler Versions}: For example, starting from Solidity version 0.4.21, the \textit{emit} keyword is introduced to trigger events, whereas earlier versions do not require it. Similarly, the \textit{receive} function becomes the standard function for receiving Ether in version 0.6.0, replacing the previous \textit{function()} syntax. 
Errors related to event triggering and the \textit{receive} function together account for 25.58\% of all compilation failures, indicating that current models have not sufficiently learned to adapt to syntax changes across different Solidity versions.

\textbf{Weak Understanding of Type System and Storage Mechanisms}: Unlike other general-purpose languages, Solidity employs a state layering mechanism, introducing \textit{storage} and \textit{memory} to represent variable locations, and variable types are limited with less flexible type conversion. The generated code frequently violates these rules, including type conversion errors (6.20\%), parameter type mismatches (4.65\%), and storage type conversion errors (1.55\%), which together account for 12,40\% of all compilation errors. These results indicate a weak understanding of Solidity’s type system and storage semantics in current models.

\textbf{Errors in Syntax Rules and Keyword Usage}: Solidity uses the \textit{pure} and \textit{view} keywords to define function visibility and \textit{override} to indicate method overriding. %
However, the generated code often misuses these keywords, with \textit{pure}/\textit{view} keyword errors accounting for 2.33\% and \textit{override} conflicts for 8.53\% of all compilation errors.
Additionally, the model exhibits flaws in handling complex Solidity-specific syntax structures, such as \textit{abi.encodePacked, abi.encodeWithSignature}, and \textit{delegatecall}. Furthermore, basic syntax errors, such as missing parentheses or semicolons (6.20\%) and function parameter count mismatches (5.43\%), are also frequent, highlighting the model's lack of proficiency in Solidity syntax compared to its performance in general-purpose languages like Python and Java.

\textbf{Insufficient Context Dependency Handling}: For contract-level code generation, understanding contextual dependencies is crucial. However, common compilation errors include duplicate variable declarations and calling undefined identifiers, both of which occur at a rate of 20.15\%. This indicates the model's insufficient grasp of contextual information within the Solidity development environment.

\begin{center}
    \begin{myboxc} \textbf{Findings 5:} In LLMs generated smart contracts, compilation errors are most commonly seen as \textit{ParseError} and \textit{TypeError}. Among them, errors caused by Solidity-specific features (such as compiler version compatibility and type system) account for 49.61\%, while errors due to insufficient context understanding account for 21.71\%.
    \end{myboxc}
\end{center}

\subsubsection{\textbf{Testing Errors}}
In our test suite, we focus on evaluating the consistency of generated contracts in three aspects: \textit{return values}, \textit{event triggers}, and \textit{storage states}. However, the generated contracts often exhibit various types of inconsistencies in these aspects. Our statistics show that return value inconsistencies have the highest proportion at 45.18\%, followed by storage state inconsistencies at 43.05\%, and event trigger inconsistencies being the lowest at 11.86\%.

These inconsistencies reveal multiple potential deficiencies in the model’s code generation capabilities. \textit{Return value inconsistencies} often result from the model’s failure to accurately understand function logic or generate erroneous code in conditional statements and loops, leading to incorrect results for callers. \textit{Event trigger inconsistencies} indicate the model’s limited understanding of the Solidity event mechanism, which may prevent critical operations from being captured in logs or blockchain explorers, reducing system transparency. \textit{Storage state inconsistencies} expose the model’s shortcomings in variable management, which may cause subsequent transactions to operate on incorrect data or even lead to security vulnerabilities, such as reentrancy attacks or unauthorized access. The following are two typical error cases:

\begin{figure}[htbp]
    \centering
    \includegraphics[page=1, width=\columnwidth, clip]{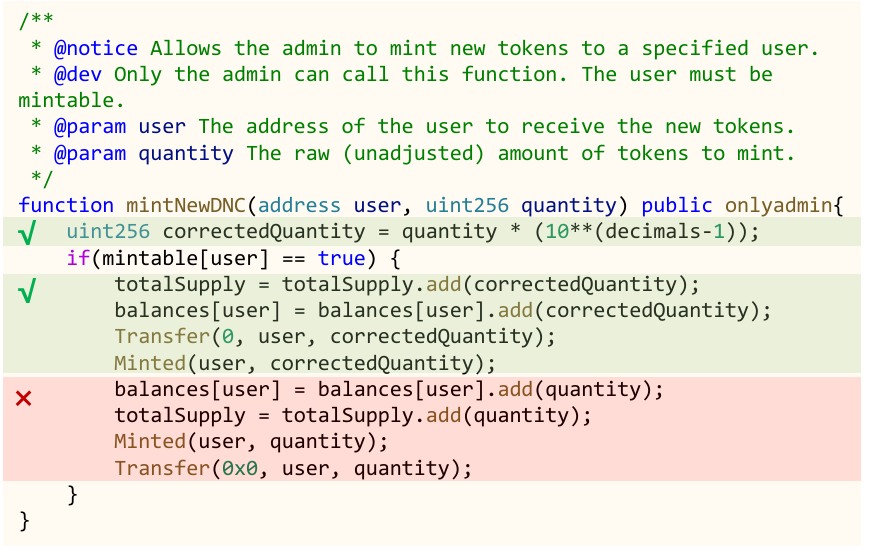}
    \caption{Example of Misunderstanding Implicit Conditions}
    \label{fig:error1}
\end{figure}

\textbf{Testing Error 1: Misunderstanding Implicit Conditions:} In the \textit{mintNewDNC} function shown in Figure~\ref{fig:error1}, the generated code fails to handle token decimal adjustments. The docstring describes parameter \textit{quantity} as a \textit{raw (unadjusted) amount}, implicitly indicating an adjustment is needed. However, the model uses the parameter directly, missing this requirement.
    
\begin{figure}[htbp]
    \centering
    \includegraphics[page=2, width=\columnwidth, clip]{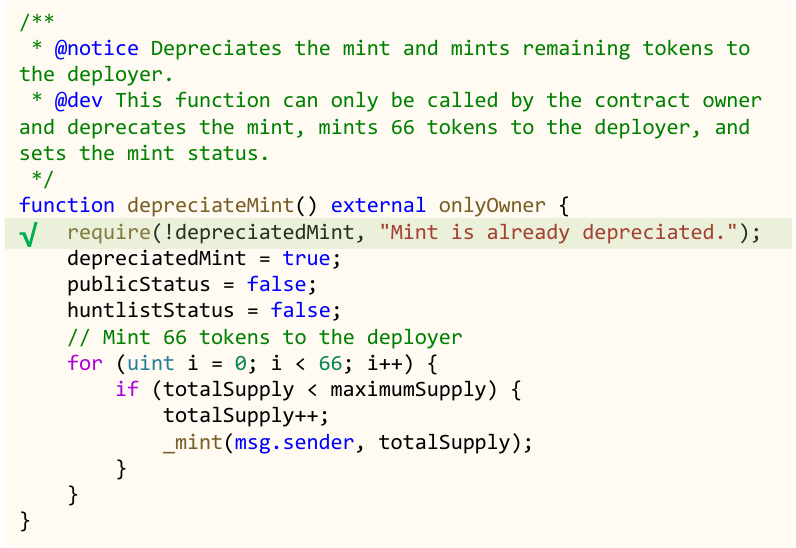}
    \caption{Example of Lack of Context Understanding}
    \label{fig:error2}
\end{figure}

\textbf{Testing Error 2: Lack of Context Understanding:} In generating the \textit{depreciateMint} function, the model fails to fully understand the context. It omits the necessary condition check for the \textit{depreciateMint} variable at the beginning of the function, which is explicitly described in comments as a flag to prevent multiple calls. This oversight allows the function to be repeatedly called, leading to functional deviations.

\section{Discussion}

\subsection{Recitation for LLMs}
\label{discussion:recitation}
In our evaluation of the Gemini-2.0-Flash-Exp model, we observe a recitation error in 12.54\% of the tests. This error occurs when the LLM detects that the generated output significantly overlaps with copyright-protected or highly similar content from the training data~\cite{GeminiAPIRecitation}. To prevent copyright infringement, the model halts generation and labels the output as ``recitation''. This mechanism serves as a safeguard against unauthorized content reproduction.

However, in the context of smart contract code generation, which is inherently highly standardized and repetitive (e.g., the ERC20 token standard), this mechanism can become problematic. Even though we have rewritten comments in our benchmark to mitigate such issues, the recurring structure and standardized nature of smart contracts make it challenging to avoid recitation. The model struggles to distinguish between legitimate reuse of standard code patterns and potential copyright infringement, resulting in false positives.

Addressing this issue is essential for future smart contract development, as it highlights the need for refined mechanisms that can accurately differentiate between valid standard code reuse and inappropriate content reproduction, thereby minimizing the risk of erroneous recitation.

\subsection{Implications}
\subsubsection{For Model Developers} 
Our study shows that current LLMs achieve insufficient performance on Solidity contract-level generation compared to function-level generation, and frequently produce \textit{ParseError} and \textit{TypeError} issues, particularly those stemming from compiler version incompatibilities, type system misuses, and insufficient handling of contextual dependencies (see Findings 2 and 5). These results indicate that existing models struggle to capture Solidity-specific syntax variations and to maintain inter-function consistency.
To address these limitations, model developers could incorporate more version-diverse training data, adopt semantics-preserving data augmentation strategies that emphasize inter-function interactions, and apply structure-aware decoding constraints to reduce syntax and type errors. Additionally, introducing compiler version information as explicit control signals during decoding may help mitigate version-related failures.

\subsubsection{For Smart Contract Developers} %
We find that while models perform well on standardized ERC20 contracts, they struggle with more complex categories such as ERC721 and DeFi, often producing logically inconsistent or incomplete implementations (see Findings 3 and 4). These categories remain challenging for current models, likely due to their intricate logic and heavy inter-contract dependencies.
Developers should therefore exercise heightened caution when adopting such code in practice. Specifically, they could conduct thorough manual reviews of dependency handling and state management logic, carefully check for subtle semantic defects such as incorrect event triggers or missing condition checks (see Section~\ref{result:case}), and avoid assuming correctness solely based on successful compilation, given the high prevalence of semantic inconsistencies revealed by our dynamic evaluation.

\subsubsection{For Researchers} 
We present SolContractEval, addressing the lack of standardized contract-level benchmarks for smart contract generation by providing a reproducible dataset built from real-world contracts with contextual integrity. Our test case construction method is generalizable and can serve as a blueprint for developing future smart contract benchmarks. Building on this foundation, future research could explore fine-grained error attribution, aiming to pinpoint errors at the level of individual components and dependencies within generated contracts, and develop hybrid evaluation approaches that combine static analysis with transaction replay to more accurately assess and localize semantic errors.

\subsection{Threat to Validity}
\subsubsection{Internal Threats} In constructing the benchmark dataset, we recognize the potential risk of data leakage between the benchmark data and the model's training data. 
To mitigate potential data leakage, two developers with three years of smart contract experience independently rewrite the original comments into structured NatSpec documentation, ensuring semantic accuracy while significantly differing from the source. All annotations are cross-verified to ensure consistency and to minimize subjective bias during manual annotation.
We also conduct a preliminary qualitative comparison with existing benchmarks from other programming domains to provide a rough performance reference(see Section \ref{result:compare}). However, due to inherent differences across programming domains, it is difficult, if not impossible, to control variables such as task design and data distribution to ensure a fully equitable comparison. Thus, our comparison should be interpreted as suggestive rather than as a rigorous quantitative evaluation of task difficulty.

\subsubsection{External Threats} While our evaluation focuses on three widely used smart contract types, which may not capture the full diversity of contract forms in the ecosystem, these types represent common real-world use cases. Moreover, SolContractEval is designed for extensibility, allowing more categories to be incorporated in future work to broaden coverage. 
Additionally, due to time and cost constraints, this study evaluates six LLMs, so findings may not fully represent all models. Nevertheless, we include representative state-of-the-art models from different families (e.g., GPT, Gemini, Claude) to ensure the reliability of our conclusions.

\section{Related Work}

\subsection{Code Generation Benchmark}
Most existing benchmarks focus on general-purpose languages~\cite{chen2021evaluating,austin2021program,yu2024codereval,du2023classeval,deng2025nocode}. For instance, HumanEval~\cite{chen2021evaluating} includes 164 tasks on language comprehension and algorithms, while MBPP~\cite{austin2021program} offers 974 beginner-level Python functions. In more complex settings, CoderEval~\cite{yu2024codereval} proposes a repository-level benchmark with 460 Python/Java tasks from open-source projects, and ClassEval~\cite{du2023classeval} introduces a class-level benchmark with 100 Python class tasks validated by test suites. Despite rapid progress in benchmarking for general-purpose languages, benchmarks tailored to Solidity remain scarce.

BenchSol~\cite{daspe2024benchmarking} is the first to evaluate LLMs on Solidity, providing 15 manually designed tasks with Hardhat tests. SolEval~\cite{peng2025soleval} extends this to a repository-level benchmark of 1,125 real-world contracts from GitHub for multi-dimensional assessment. SolBench~\cite{chen2025solbench} further emphasizes function-level generation, collecting 4,178 functions from on-chain contracts for functional testing.
These benchmarks have several limitations: 1) they primarily focus on function-level generation or overly simplistic contracts, failing to capture the complexity of real-world scenarios; 2) functional testing is often conducted using manually constructed cases or fuzzing, which makes it hard to capture issues that may arise in real applications.

In contrast, our \textbf{\textit{SolContractEval}} is the first contract-level benchmark for Solidity code generation, differing from prior work in both task design and evaluation. Built from real deployed contracts, it requires models to generate full contracts with contextual dependencies such as inheritance and inter-function interactions. To assess functional correctness, SolContractEval replays historical on-chain transactions and analyzes runtime behavior from multiple perspectives.

\subsection{Code Generation in Smart Contract}

Dade et al.\cite{dade2023optimizing} proposed MazzumaGPT, an optimized LLM for Web3 smart contract generation, enhancing efficiency while preserving functional correctness. Wang et al.\cite{zhang2025codebc} introduced CodeBC, which uses a three-stage fine-tuning strategy to optimize CodeLlama. By leveraging a security tag mechanism to guide generation, it mitigates security risks without manual vulnerability labels. Napoli et al.~\cite{napoli2024leveraging} developed a contract generation system based on the CO-STAR prompt optimization framework, integrating tools like Slither to form a comprehensive quality assurance pipeline.
These works remain in the exploratory stage, facing challenges such as inconsistent evaluation standards and limited testing scenarios. Our work complements them by providing a comprehensive and objective benchmark through systematic test case design and multi-dimensional evaluation metrics.

\section{Conclusion}

In this paper, we propose \textbf{\textit{SolContractEval}}, a contract-level benchmark for Solidity code generation, addressing the current lack of evaluation tools for LLMs in real-world Solidity development scenarios. We conduct the first empirical study on contract-level code generation using six state-of-the-art LLMs. Our results show that Claude-3.7-Sonnet and o4-mini lead in contract-level Solidity generation performance. Second, LLMs underperform compared to class-level generation in general-purpose tasks. Notably, ERC20 contracts, the most widely used category, also yield the most stable generation results. Furthermore, Solidity’s inherent characteristics, such as its strict type system, compiler version dependencies, and layered storage model, pose major challenges for accurate contract generation.

\section*{Acknowledgment}
This research/project was supported by the National Natural Science Foundation of China (No.62202420), Zhejiang Provincial Natural Science Foundation of China (No.LZ25F020003), the National Natural Science Foundation of China (No.62372398 and No.72342025), and the Fundamental Research Funds for the Central Universities (No.226-2025-00067).

\section*{Data Availability}
All code and data for this study are publicly available at https://github.com/ZJU-CTAG/SolContractEval.

\balance
\bibliographystyle{IEEEtran}
\bibliography{reference}

\end{document}